\documentclass[twocolumn,showpacs,preprintnumbers,amsmath,amssymb,superscriptaddress]{revtex4}

\usepackage{graphicx}  % Include figure files
\usepackage{dcolumn}   % Align table columns on decimal point
\usepackage{bm}        % bold math

\newcommand{\met}{\,/\!\!\!\!E_{T}}
\newcommand{\et}{E_T}

\newcommand{\ppbar}{{p\bar{p}}}

\newcommand{\zprime}{Z^\prime}
\newcommand{\aff}[1]{\ignorespaces $^{#1}$}

\begin{document}

%\preprint{hep-ex/05xxxxx}

\title{Search for New Physics Using High Mass Tau Pairs from
      1.96-TeV $p\bar{p}$ Collisions}%

\date{June 13, 2005}

\author{
D.~Acosta,\aff{16} J.~Adelman,\aff{12} T.~Affolder,\aff{9} T.~Akimoto,\aff{54}
M.G.~Albrow,\aff{15} D.~Ambrose,\aff{15} S.~Amerio,\aff{42}
D.~Amidei,\aff{33} A.~Anastassov,\aff{50} K.~Anikeev,\aff{15}
% ...
}
\affiliation{
\aff{1}  {\small\it Institute of Physics, Academia Sinica, Taipei, Taiwan 11529,
Republic of China}, \\
\aff{2}  {\small\it Argonne National Laboratory, Argonne, Illinois 60439}, \\
\aff{3}  {\small\it Institut de Fisica d'Altes Energies, Universitat Autonoma
de Barcelona, E-08193, Bellaterra (Barcelona), Spain}, \\
\aff{4}  {\small\it Istituto Nazionale di Fisica Nucleare, University of Bologna,
I-40127 Bologna, Italy}, \\
\aff{5}  {\small\it Brandeis University, Waltham, Massachusetts 02254}, \\
\aff{6}  {\small\it University of California, Davis, Davis, California  95616}, \\
\aff{7}  {\small\it University of California, Los Angeles, Los
Angeles, California  90024}, \\
\aff{8}  {\small\it University of California, San Diego, La Jolla, California  92093}, \\
\aff{9}  {\small\it University of California, Santa Barbara, Santa Barbara, California
93106}, \\
\aff{10} {\small\it Instituto de Fisica de Cantabria, CSIC-University of Cantabria,
39005 Santander, Spain}, \\
\aff{11} {\small\it Carnegie Mellon University, Pittsburgh, PA  15213}, \\
\aff{12} {\small\it Enrico Fermi Institute, University of Chicago, Chicago,
Illinois 60637}, \\
\aff{13}  {\small\it Joint Institute for Nuclear Research, RU-141980 Dubna, 
Russia}, \\ 
\aff{14} {\small\it Duke University, Durham, North Carolina  27708}, \\
\aff{15} {\small\it Fermi National Accelerator Laboratory, Batavia, Illinois
60510}, \\
\aff{16} {\small\it University of Florida, Gainesville, Florida  32611}, \\
\aff{17} {\small\it Laboratori Nazionali di Frascati, Istituto Nazionale di Fisica
               Nucleare, I-00044 Frascati, Italy}, \\
\aff{18} {\small\it University of Geneva, CH-1211 Geneva 4, Switzerland}, \\
\aff{19} {\small\it Glasgow University, Glasgow G12 8QQ, United Kingdom}, \\
\aff{20} {\small\it Harvard University, Cambridge, Massachusetts 02138}, \\
\aff{21} {\small\it Division of High Energy Physics, Department of
Physics, University of Helsinki and Helsinki Institute of Physics,
FIN-00014, Helsinki, Finland}, \\
\aff{22} {\small\it Hiroshima University, Higashi-Hiroshima 724, Japan}, \\
\aff{23} {\small\it University of Illinois, Urbana, Illinois 61801}, \\
\aff{24} {\small\it The Johns Hopkins University, Baltimore, Maryland 21218}, \\
\aff{25} {\small\it Institut f\"{u}r Experimentelle Kernphysik,
Universit\"{a}t Karlsruhe, 76128 Karlsruhe, Germany}, \\
\aff{26} {\small\it High Energy Accelerator Research Organization (KEK), Tsukuba,
Ibaraki 305, Japan}, \\
\aff{27} {\small\it Center for High Energy Physics: Kyungpook National
University, Taegu 702-701; Seoul National University, Seoul 151-742; and
SungKyunKwan University, Suwon 440-746; Korea}, \\
\aff{28} {\small\it Ernest Orlando Lawrence Berkeley National Laboratory,
Berkeley, California 94720}, \\
\aff{29} {\small\it University of Liverpool, Liverpool L69 7ZE, United Kingdom}, \\
\aff{30} {\small\it University College London, London WC1E 6BT, United Kingdom}, \\
\aff{31} {\small\it Massachusetts Institute of Technology, Cambridge,
Massachusetts  02139}, \\
\aff{32} {\small\it Institute of Particle Physics: McGill University,
Montr\'{e}al, Canada H3A~2T8; and University of Toronto, Toronto, Canada
M5S~1A7}, \\
\aff{33} {\small\it University of Michigan, Ann Arbor, Michigan 48109}, \\
\aff{34} {\small\it Michigan State University, East Lansing, Michigan  48824}, \\
\aff{35} {\small\it Institution for Theoretical and Experimental Physics, ITEP,
Moscow 117259, Russia}, \\
\aff{36} {\small\it University of New Mexico, Albuquerque, New Mexico 87131}, \\
\aff{37} {\small\it Northwestern University, Evanston, Illinois  60208}, \\
\aff{38} {\small\it The Ohio State University, Columbus, Ohio  43210}, \\
\aff{39} {\small\it Okayama University, Okayama 700-8530, Japan}, \\
\aff{40} {\small\it Osaka City University, Osaka 588, Japan}, \\
\aff{41} {\small\it University of Oxford, Oxford OX1 3RH, United Kingdom}, \\
\aff{42} {\small\it University of Padova, Istituto Nazionale di Fisica
          Nucleare, Sezione di Padova-Trento, I-35131 Padova, Italy}, \\
\aff{43} {\small\it University of Pennsylvania, Philadelphia,
        Pennsylvania 19104}, \\
\aff{44} {\small\it Istituto Nazionale di Fisica Nucleare Pisa, Universities 
of Pisa, Siena and Scuola Normale Superiore, I-56127 Pisa, Italy}, \\
\aff{45} {\small\it University of Pittsburgh, Pittsburgh, Pennsylvania 15260}, \\
\aff{46} {\small\it Purdue University, West Lafayette, Indiana 47907}, \\
\aff{47} {\small\it University of Rochester, Rochester, New York 14627}, \\
\aff{48} {\small\it The Rockefeller University, New York, New York 10021}, \\
\aff{49} {\small\it Istituto Nazionale di Fisica Nucleare, Sezione di Roma 1,
University di Roma ``La Sapienza," I-00185 Roma, Italy}, \\
\aff{50} {\small\it Rutgers University, Piscataway, New Jersey 08855}, \\
\aff{51} {\small\it Texas A\&M University, College Station, Texas 77843}, \\
\aff{52} {\small\it Texas Tech University, Lubbock, Texas 79409}, \\
\aff{53} {\small\it Istituto Nazionale di Fisica Nucleare, University of Trieste/\
Udine, Italy}, \\
\aff{54} {\small\it University of Tsukuba, Tsukuba, Ibaraki 305, Japan}, \\
\aff{55} {\small\it Tufts University, Medford, Massachusetts 02155}, \\
\aff{56} {\small\it Waseda University, Tokyo 169, Japan}, \\
\aff{57} {\small\it Wayne State University, Detroit, Michigan  48201}, \\
\aff{58} {\small\it University of Wisconsin, Madison, Wisconsin 53706}, \\
\aff{59} {\small\it Yale University, New Haven, Connecticut 06520}}
\collaboration{CDF Collaboration}

\begin{abstract}

We present the results of a search for anomalous resonant production
of tau lepton pairs with large invariant mass, the first such search
using the CDF~II Detector in Run~II of the Tevatron $p\bar{p}$
collider.  Such anomalous production could arise from various new
physics processes.  In a data sample corresponding to 195~pb$^{-1}$
of integrated luminosity we predict 2.8$\pm$0.5 events from Standard
Model background processes and observe 4.  We use this result to set
limits on the production of heavy scalar and vector particles decaying
to tau lepton pairs.

\end{abstract}

%\keywords{Suggested keywords} % Use showkeys class option if keyword
                               % display desired
\maketitle

%------------------------------------------------------------------

% Introduction ---------------------------------------------------------

At the Fermilab Tevatron $p\bar{p}$ collider, a number of
non--Standard-Model physics processes can lead to events with
high-mass tau lepton pairs in the final state.  Examples include the
resonant production of Higgs scalars in two-Higgs-doublet
models~\cite{ref-twohiggs} at large $\tan\beta$, the ratio of the
vacuum expectation value of the two doublets.  Two Higgs doublets are
required, for example, in the Minimal Supersymmetric Standard
Model~\cite{ref-mssm}, a favored candidate for extending the Standard
Model.  The heavy scalar and pseudoscalar Higgs bosons in this theory
would decay to tau pairs about 9\% of the time.  Also, in
supersymmetry, if R-parity is not conserved, heavy scalar neutrino
production could have tau pair decay modes~\cite{ref-rpvsnu}.  If
there are heavy $Z^\prime$ bosons, these could also produce high mass
tau pairs in the final state, possibly even with enhanced tau
couplings~\cite{ref-zprime}.  With the large new data sample from
Run~II of the Tevatron it is thus of great interest to perform a
generic search for high-mass tau pairs.

This Letter presents the results of a search for high-mass tau pairs
performed using CDF~II, the upgraded Collider Detector at Fermilab
(CDF)~\cite{ref-thesis}.  In 2002 and 2003 CDF recorded a data sample
corresponding to 195~pb$^{-1}$ of integrated luminosity of $p\bar{p}$
collisions at a center of mass energy of 1.96~TeV.  This is the first
such search with the new high-statistics data sample~\cite{ref-amy}
and new tau identification techniques.

% Overview of method --------------------------------------------------

Since the tau lepton decays to lighter leptons ($e$ or $\mu$) about
35\% of the time, and to low-multiplicity hadronic states the rest of
the time, this analysis selects events with one identified hadronic
tau decay ($\tau_h$) and one other tau decay in the final state.
Thus, there are three distinct final states, which we denote
$e\tau_h$, $\mu\tau_h$, and $\tau_h\tau_h$.  The main background to
this search comes from Drell-Yan (DY) $Z/\gamma^*\to\tau^+\tau^-$
production.  Since we seek new particles with mass much larger than
that of the $Z$, we use the observed rate for this background (at
smaller tau pair masses) as a control sample, and define the signal
region as that where the tau pairs have large visible invariant mass,
with missing energy due to the neutrinos from the tau decays.

% CDF Detector ---------------------------------------------------------

CDF II is a large general purpose detector with an overall cylindrical
geometry surrounding the $p\bar{p}$ interaction region~\cite{ref-cdf}.
The three-dimensional trajectories of charged particles produced in
$p\bar{p}$ collisions are measured starting at radii of 1.5~cm with
multiple layers of silicon microstrip detectors, and are measured at
outer radii with an axial/stereo wire drift chamber (COT).  The
tracking system lies inside a uniform 1.4-T magnetic field produced by
a superconducting solenoid, with the field oriented along the beam
direction.  Outside the solenoid lie the electromagnetic calorimeter
and the hadronic calorimeters, which are segmented in pseudorapidity
($\eta$)~\cite{ref-eta} and azimuth in a projective ``tower''
geometry.  A set of strip/wire chambers (CES) located at a depth of
six radiation lengths aids in reconstructing photons and electrons
from the shower shape.  Muons are identified by a system of drift
chambers placed outside the calorimeter steel, which acts as an
absorber for hadrons.  The integrated luminosity of the $\ppbar$
collisions is measured to an accuracy of 6\% using the Cerenkov
Luminosity Counters~\cite{ref-lum}.

% Trigger ---------------------------------------------------------------

The $e\tau_h$ and $\mu\tau_h$ events of interest are recorded using
triggers designed to select ``lepton plus track'' events: those with
an $e$ or $\mu$ with transverse momentum ($p_T$) greater than
8~GeV/$c$ and another charged track with $p_T > 5$ GeV/$c$ identified
by the eXtremely Fast Tracker (XFT) portion of the trigger
electronics~\cite{ref-xft} which reconstructs charged tracks in the
COT.  The efficiency of this trigger is measured using leptons from
$Z$ boson decays, the $\Upsilon$ resonance, and photon
conversions~\cite{ref-NIM}.

For selecting $\tau_h\tau_h$ events we use a trigger designed to
select at least one hadronically decaying tau with $E_T > 20$ GeV
accompanied by at least 25~GeV missing energy in the plane transverse
to the beam direction ($\met$).  The tau is identified by matching an
XFT track with $p_T > 5$ GeV/$c$ to a calorimeter cluster.  Data used
from this trigger come from a sample corresponding to the first
72~pb$^{-1}$ of integrated luminosity recorded; this is less than that
of the rest of the data used because of subsequent changes due to rate
limitations.

% Event Selection -------------------------------------------------------

Events selected by the triggers were recorded and processed later to
reconstruct charged particle tracks, calorimeter clusters, and to
identify electrons, muons, photons, jets, and $\met$.  Electrons and
muons are reconstructed using algorithms described in
Ref.~\cite{ref-cdf}.  Identification of hadronic decays of taus
employs a novel ``shrinking cone'' algorithm based on high-$p_T$
charged tracks in the silicon/COT system, and $\pi^0$ candidates
identified using the CES by matching strip clusters with wire clusters
based on the energy in each.

\begin{figure}[t]
  \begin{center}
    \includegraphics[width=\columnwidth]{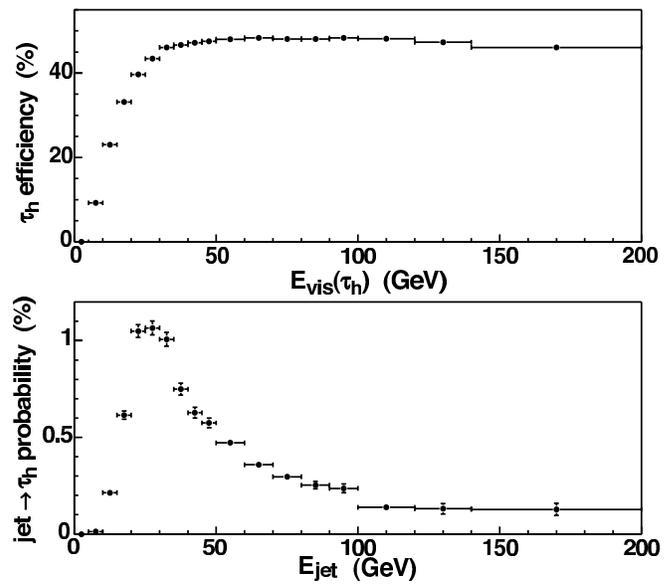}
  \end{center}
  \caption{Identification efficiency (top) and jet$\to\tau_h$ probability 
           (bottom) as a function of generated tau visible decay product 
           energy and measured jet energy, respectively.  The top plot, 
           based on simulation,shows the probability that true hadronically 
           decaying taus are identified as $\tau_h$ using the selection in 
           the text.  The lower plot  shows the probability that hadronic 
           jets in the range $|\eta|<1$ are misidentified as $\tau_h$, as 
           a function of jet calorimeter cluster energy.  The error bars
           indicate the statistical uncertainties.}
  \label{fig-fake}
\end{figure}

The $\tau_h$ identification algorithm begins with a list of ``seed
tracks'' ranked in $p_T$, not yet used for another tau candidate, and
having $p_T > 6$ GeV/$c$.  Then it finds the number of other tracks
with $p_T > 1$ GeV/$c$ and $\pi^0$ candidates with at least 1 GeV
whose momentum vector makes an angle of less than $\alpha$ with the
seed track.  The angle $\alpha$ is a function of $E_{clu}$, the energy
in the calorimeter cluster associated with the seed track.  The value
of $\alpha$ is 10$^\circ$ or (5 GeV)/$E_{clu}$ radians, whichever is
less.  To allow for resolution effects, the value of $\alpha$ is not
less than 100 mrad for $\pi^0$ candidates, or 50 mrad for charged
tracks.  If any other tracks or $\pi^0$ candidates have an angle
greater than $\alpha$ but less than 30$^\circ$ to the seed track, or
if the invariant mass calculated from the sum of all charged track and
$\pi^0$ candidate four-momenta exceeds 1.8~GeV/$c^2$, the $\tau_h$
candidate is rejected.  Also, if there is at least 2 GeV of
electromagnetic energy in calorimeter towers not part of the tau
cluster, and whose centers have $\Delta R = \sqrt{(\Delta\eta)^2 +
(\Delta\phi)^2} < 0.4$ from the tau seed track direction, the tau
candidate is rejected.  Candidates with momentum having an angle of
less than 10$^\circ$ with that of of a previously identified $e$ or
$\mu$ are rejected. The algorithm then considers further possible seed
tracks, repeating the process until none remain.

The main challenge comes from the large production rate of hadronic
jets, which can be misidentified as $\tau_h$.  Using the selection
described above, Figure~\ref{fig-fake} shows the efficiency for real
hadronically decaying taus with $|\eta|<1$ to be reconstructed as
$\tau_h$, using the simulation discussed below.  The figure also shows
the jet $\to \tau_h$ ``fake'' probability that hadronic jets are
misidentified as hadronic tau decays.  These jets, reconstructed in a
cone size of $\Delta R = 0.7$, come from events recorded with triggers
requiring various thresholds for calorimeter cluster energy.

To discriminate against background, for the $e\tau_h$ ($\mu\tau_h$)
channel the electron (muon) must have a transverse energy of at least
10 GeV, the $\tau_h$ must have $\et > 25$ GeV, and the event must have
$\met >$15 GeV.  For the $\tau_h\tau_h$ channel, one $\tau_h$ must
have $\et$ greater than 25 GeV, and the other must have at least 10~
GeV.  The azimuthal angle between the $\met$ vector and the $e$ or
$\mu$ (in $e\tau_h$ or $\mu\tau_h$ events) or the less-energetic of
the two in $\tau_h\tau_h$ events must be less than 30$^\circ$.

% Visible mass: signal and control --------------------------------------

For all events selected by the above cuts we calculate the ``visible
mass'' ($m_{vis}$) by adding the measured four-momenta of the two
identified tau decay products in the event to the missing transverse
energy four-momentum (for which the $z$ component is taken as zero),
and then calculating the invariant mass of the sum.  This quantity
efficiently distinguishes between lower-mass production of tau pairs
(mainly from $Z$ boson decays) and high-mass tau pairs from possible
new massive resonant particle production.

% Backgrounds -----------------------------------------------------------

The main source of events expected in the selected sample is DY
production of $Z/\gamma^*$ decaying to lepton pair final states, and
of these, tau pair production predominates.  The production cross
section times branching ratio to pairs of each charged lepton species
for DY $Z/\gamma^*$ is assumed to be 250~pb~\cite{ref-zxsec} in the
mass range 66-116 GeV/$c^2$.  For the DY process and for the possible
new physics processes discussed below, we simulate the production and
decay using the PYTHIA 6.215 Monte Carlo program~\cite{ref-pythia}
with CTEQ5L parton distribution functions (PDF's)~\cite{ref-CTEQ5L},
with tau decays simulated by TAUOLA~\cite{ref-tauola}.  Acceptance and
resolution effects come from the full CDF~II detector simulation.

The second largest source of events passing our selection criteria is
hadronic jets which are misidentified as a $\tau_h$, for example from
events with a $W$ boson decaying to a charged lepton and a neutrino
plus a jet which passes the $\tau_h$ identification criteria. The
estimated number of expected events comes from applying the jet $\to
\tau_h$ ``fake'' rates to jets in events passing the trigger and other
requirements, excluding the $\tau_h$ identification.

% Systematic Errors -----------------------------------------------------

Various systematic uncertainties affect the predicted number of signal
and background events.  The largest is due to imperfect modeling of
the tau identification efficiency.  We perform a cross check of this
efficiency using $W\to\tau\nu$ events recorded in the first
72~pb$^{-1}$.  Assuming a production cross section times branching
ratio to $\tau\nu$ of 2688 pb~\cite{ref-zxsec}, this check yields a
multiplicative factor of 0.97$\pm$0.10, which is incorporated into the
acceptance calculation in the simulation.  The 10\% uncertainty in
this factor, which affects each identified $\tau_h$ in the selected
sample, includes the trigger efficiency uncertainty.

The uncertainties in the $e$ and $\mu$ identification and trigger
efficiency, of 4\% for $e$ and 5.5\% for $\mu$, come from studies
described elsewhere~\cite{ref-thesis}.

The jet $\to\tau_h$ fake background estimate has a 20\% uncertainty
reflecting the variation in the fake rate among the different
trigger samples.

A 6\% uncertainty due to imperfect modeling of the $\met$ comes from 
studies of transverse energy balancing in events with high energy jets 
recoiling against high energy photons. 

Imperfect knowledge of the PDF's leads to an 8\% uncertainty in the DY
and any new physics signal acceptances.  The uncertainty is estimated
from the variation of the acceptance using different PDF sets.

\begin{table}
  \caption{Mean expected numbers of events in the control region
           ($m_{vis}<120$ GeV/$c^2$), compared with the numbers observed.  
           The uncertainties listed include both statistical and systematic effects.}
  \begin{tabular}{ccccc} \hline \hline
   source                   & $e\tau_h$   & $\mu\tau_h$ &$\tau_h\tau_h$&   total     \\ \hline
 $Z/\gamma^*\to e^+ e^-$    & 0.1$\pm$0.1 &     -       &     -       & 0.1$\pm$0.1  \\
 $Z/\gamma^*\to\mu^+\mu^-$  &     -       & 0.5$\pm$0.3 &     -       & 0.5$\pm$0.3  \\
 $Z/\gamma^*\to\tau^+\tau^-$& 45$\pm$7    &  38$\pm$6   & 4.2$\pm$0.8 & 88$\pm$12    \\
 jet $\to\tau_h$            &  4$\pm$1    &   4$\pm$1   & 3.2$\pm$0.6 & 11$\pm$2     \\ \hline
   total expected           & 49$\pm$7    &  43$\pm$6   & 7.4$\pm$1.0 & 99$\pm$13    \\ \hline
   observed                 &     46      &    36       &     8       &    90        \\ \hline \hline
  \end{tabular}
  \label{tab-control}
\end{table}

\begin{table}
  \caption{Mean expected numbers of events in the signal region
           ($m_{vis}>120$ GeV/$c^2$), compared with the numbers observed. (The
           numbers shown are rounded after adding.)
           The uncertainties listed include both statistical and systematic effects.}
  \begin{tabular}{ccccc} \hline \hline
   source                   & $e\tau_h$   & $\mu\tau_h$ &$\tau_h\tau_h$&   total     \\ \hline
 $Z/\gamma^*\to e^+ e^-$    & 0.2$\pm$0.1 &     -       &     -       & 0.2$\pm$0.1 \\
 $Z/\gamma^*\to\mu^+\mu^-$  &     -       & 0.5$\pm$0.3 &     -       & 0.5$\pm$0.3 \\
 $Z/\gamma^*\to\tau^+\tau^-$& 0.6$\pm$0.1 & 0.5$\pm$0.1 & 0.4$\pm$0.1 & 1.4$\pm$0.3 \\
 jet $\to\tau_h$            & 0.3$\pm$0.1 & 0.2$\pm$0.1 & 0.3$\pm$0.1 & 0.8$\pm$0.2 \\ \hline
   total expected           & 1.0$\pm$0.2 & 1.2$\pm$0.3 & 0.6$\pm$0.1 & 2.8$\pm$0.5 \\ \hline
   observed                 &     4       &     0       &     0       &    4        \\ \hline \hline
  \end{tabular}
  \label{tab-signal}
\end{table}

% Control Results -------------------------------------------------------

Table~\ref{tab-control} summarizes the expected numbers of events by
source for each channel, and shows the observed number of events in
each search channel, for the control region dominated by $Z$ boson
decay ($m_{vis}<120$ GeV$/c^2$).  The observed number is in good
agreement with that expected.  This gives confidence that the
estimated efficiencies and background rates are well understood, and
we proceed to examine the signal region.

% Signal Results --------------------------------------------------------

Table~\ref{tab-signal} shows, for each search channel, the numbers of
events expected and the uncertainty for each background source in the
signal region ($m_{vis}> 120$ GeV/$c^2$).  We observe four $e\tau_h$
events, and no $\mu\tau_h$ or $\tau_h\tau_h$ events.  Given the 
uncertainties shown in the table, the observed number of events is in
good agreement with that expected.

\begin{figure}
  \begin{center}
    \includegraphics[width=\columnwidth]{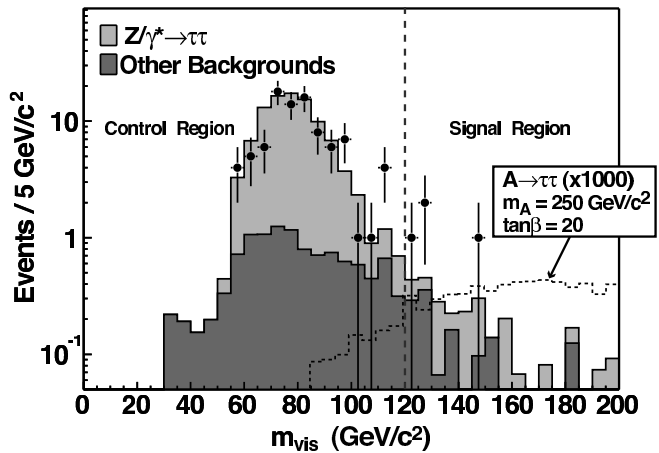}
  \end{center}
  \caption{Distribution of visible mass ($m_{vis}$) for data (points) 
           and predicted backgrounds (shaded histograms) in the signal and
           control regions.  The dashed histogram shows the distribution 
           expected for a pseudoscalar Higgs $A$, with $m_A = 250$ GeV/$c^2$
           and $\tan\beta=20$, with the normalization increased by 1000.
           There are no observed events with $m_{vis}>$200~GeV/$c^2$.}
  \label{fig-mvis}
\end{figure}

Figure~\ref{fig-mvis} shows the distribution of visible mass in the 
signal and control regions, for the observed events and the predicted
background.  The distribution of the masses of the four events in the 
signal region is consistent with that expected from background.

% Limits ----------------------------------------------------------------

\begin{figure}[t!]
  \begin{center}
    \includegraphics[width=\columnwidth]{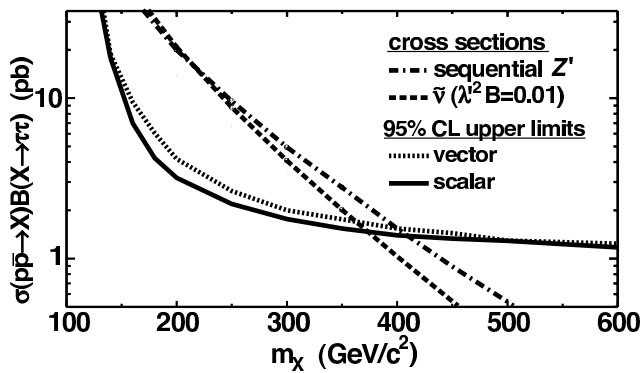}
  \end{center}
  \caption{\label{fig-limits}
           Upper limits at 95\% CL on the production cross section times 
           branching ratio to tau pairs of scalar and vector particles, 
           as a function of particle mass.  The figure also shows the 
           cross section times tau pair branching ratio for scalar neutrinos
           and sequential $Z^\prime$ bosons.}
\end{figure}

Since we observe no significant excess rate of high-mass tau pair
production, we determine upper bounds on the production cross section
times branching ratio to tau pairs of hypothetical scalar and vector
particles.  As a general model for the acceptance for scalar particle
production we use pseudoscalar Higgs boson ($A$) production, and for
vector particle production we use a $\zprime$ boson.  The acceptance
for both increases from near zero at masses of 100 GeV/$c^2$ to about
4\% at high masses (500 GeV/$c^2$ or more).

To determine the upper bounds on the cross section times branching
ratio we form a likelihood from the joint Poisson probability of all
search channel results, and use a Bayesian method to incorporate the
effects of systematic uncertainties, which are represented by
truncated gaussian prior probability densities, including
correlations.  The likelihood is converted to a posterior probability
density in the signal cross section using Bayes Theorem, assuming a
prior in the signal rate which is uniform up to some high cutoff.  The
95\% CL upper limit comes from the integral of the posterior density.

Figure~\ref{fig-limits} shows the 95\% CL upper bound on the cross
section times branching ratio to tau pairs for scalar and vector
particle production.  Table~\ref{tab-result} lists the upper limits on
the production rate of scalar and vector particles as a function of
mass.  As an example of the sensitivity, these results would rule out
a $Z^\prime$ with Standard Model couplings having a mass less than
399~GeV/$c^2$, as indicated by the curve of cross section times
branching ratio in the figure.  The figure also shows the case of
R-parity-violating scalar neutrino production and decay to tau pairs;
this analysis, as an example, excludes a 377 GeV/$c^2$ scalar 
neutrino having coupling $\lambda^\prime$ to $d\bar{d}$ and branching
ratio $B$ to tau pairs such that $\lambda^{\prime 2}B=$0.01.  In
general the limits are readily interpreted within the context of new
physics models in which new scalar or vector particles decay to tau
lepton pairs.

\begin{table*}
  \caption{The 95\% CL upper limits on scalar and vector
           particle production and decay to tau pairs.}
\begin{tabular}{lcccccccccccc}\hline\hline
mass (GeV/$c^2$)  & 120  & 140  & 160  & 180  & 200  & 250  & 300  & 350  & 400  & 450  & 500  & 600  \\ \hline
scalar limit (pb) & 87.3 & 17.9 & 7.00 & 4.23 & 3.19 & 2.19 & 1.76 & 1.54 & 1.40 & 1.33 & 1.29 & 1.17 \\ 
vector limit (pb) & 122  & 18.9 & 9.45 & 6.07 & 4.19 & 2.64 & 2.00 & 1.76 & 1.54 & 1.44 & 1.30 & 1.24 \\ \hline \hline
\end{tabular}
  \label{tab-result}
\end{table*}

% Acknowledgements-------------------------------------------------------

We thank the Fermilab staff and the technical staffs of the
participating institutions for their vital contributions. This work
was supported by the U.S. Department of Energy and National Science
Foundation; the Italian Istituto Nazionale di Fisica Nucleare; the
Ministry of Education, Culture, Sports, Science and Technology of
Japan; the Natural Sciences and Engineering Research Council of
Canada; the National Science Council of the Republic of China; the
Swiss National Science Foundation; the A.P. Sloan Foundation; the
Bundesministerium f\"ur Bildung und Forschung, Germany; the Korean
Science and Engineering Foundation and the Korean Research Foundation;
the Particle Physics and Astronomy Research Council and the Royal
Society, UK; the Russian Foundation for Basic Research; the Comisi\'on
Interministerial de Ciencia y Tecnolog\'{\i}a, Spain; in part by the
European Community's Human Potential Programme; %under contractHPRN-CT-2002-00292; 
and the Academy of Finland.

% Bibliography ----------------------------------------------------------


\begin{thebibliography}{99}
\bibitem{ref-eta}     {We define the pseudorapidity $\eta$ of a particle
                         three-momentum as $\eta \equiv -\ln(\tan\frac{\theta}{2})$, where
                         $\theta$ is the polar angle of the momentum to the beam direction.}
\end{thebibliography}
\end{document}